\documentclass[final]{cvpr}

\usepackage{times}
\usepackage{epsfig}
\usepackage{graphicx}
\usepackage{amsmath}
\usepackage{amssymb}
\usepackage{appendix}
\usepackage{booktabs}
\usepackage{pifont}
\newcommand{\cmark}{\ding{51}}
\newcommand{\xmark}{\ding{55}}

\usepackage[pagebackref=true,breaklinks=true,colorlinks,bookmarks=false]{hyperref}

\begin{document}

\title{How to Exploit the Transferability of\\Learned Image Compression to Conventional Codecs}

\author{Jan P. Klopp\\
National Taiwan University\\
{\tt\small kloppjp@gmail.com}
\and
Keng-Chi Liu\\
Taiwan AI Labs\\
{\tt\small calvin89029@gmail.com}
\and
Liang-Gee Chen, Shao-Yi Chien\\
National Taiwan University\\
{\tt\small \{lgchen,sychien\}@ntu.edu.tw}
}
\date{}
\maketitle

\begin{abstract}
   Lossy image compression is often limited by the simplicity of the chosen loss measure. Recent research suggests that generative adversarial networks have the ability to overcome this limitation and serve as a multi-modal loss, especially for textures. Together with learned image compression, these two techniques can be used to great effect when relaxing the commonly employed tight measures of distortion. However, convolutional neural network based algorithms have a large computational footprint. Ideally, an existing conventional codec should stay in place, which would ensure faster adoption and adhering to a balanced computational envelope.
   
   As a possible avenue to this goal, in this work, we propose and investigate how learned image coding can be used as a surrogate to optimize an image for encoding. The image is altered by a learned filter to optimise for a different performance measure or a particular task. Extending this idea with a generative adversarial network, we show how entire textures are replaced by ones that are less costly to encode but preserve sense of detail. 
   
   Our approach can remodel a conventional codec to adjust for the MS-SSIM distortion with over 20\% rate improvement without any decoding overhead. On task-aware image compression, we perform favourably against a similar but codec-specific approach. 
\end{abstract}


\section{Introduction}

Forming the basis for video compression, image compression's efficiency is vital to many data transmission and storage applications. 
Most compression algorithms are lossy, i.e. they do not reproduce the original content exactly but allow for deviations that reduce the coding rate. Lossy compression optimises the objective
\begin{equation}\label{eq:rate_distortion}
\mathcal{L}=R+\lambda D
\end{equation}
where $R$ and $D$ stand for rate and distortion, respectively, and $\lambda$ controls their weight relative to each other. In practice, computational efficiency is another constraint as at least the decoder needs to be able to process high resolutions in real-time under a limited power envelope, typically necessitating dedicated hardware implementations. Requirements for the encoder are more relaxed, often allowing even offline encoding without demanding real-time capability.

Recent research has developed along two lines: evolution of exiting coding technologies, such as H264 \cite{Wiegand2003} or H265 \cite{Sullivan2012a}, culminating in the most recent AV1 codec, on the one hand. On the other hand, inspired by the success of deep learning in computer vision, based on the variational autoencoder \cite{Kingma2014}, several approaches for learned image compression were developed. Through careful modelling of the latent code's symbol probabilities, those learned codecs have recently shown to outperform the x265 codec on various performance measures \cite{Minnen2018}. This trades engineering effort for complexity: training a neural network as image codec is simple compared to designing a codec with all its bells and whistles, however, the resulting neural decoder requires on the order of $10^5$ operations per pixel, where a conventional codec has at most a few hundred operations, being several orders of magnitude more efficient, an important factor given how much video and imagery is consumed on mobile devices.

Another advantage of learned compression is the ability to be easily adopted to different distributions or distortion measures, though. By merely collecting data and choosing an appropriate distortion loss, a new codec can quickly be trained without re-engineering coding tools or internal parameters of conventional compression codecs manually. 

In this work, we are taking a step towards leveraging the adaptability from learned codecs for conventional codecs without changing the conventional codec itself. In doing so, we enable a higher compression performance without touching the decoder, meaning that existing end point decoder implementations do not require changes. In lossy compression, adaptability has two aspects: dropping information that is hard to encode and of little importance and, secondly, exploiting expensive redundancies in the data. The latter would require either reorganising data or changing the coding tools a codec can apply at encoding time. The former, though, is feasible as the data can be changed prior to letting the encoder map it to its coded representation. 

We demonstrate three possible applications of this scheme:
\begin{enumerate}
	\item We apply a filtering mechanism to siphon out information that would drive the coding cost but is only of little concern with respect to the chosen loss function. Introducing a learned codec as a surrogate gradient provider conventional codecs can be refitted to the MS-SSIM distortion measure at over 20\% rate improvement.
	\item We apply the approach to task-specific compression where we outperform a competing approach on image classification without having to model the target codec specifically.
	\item We pair the surrogate-induced filter with a generative adversarial network to alter the image beyond optimization for simple distortion measures. Textures are replaced with perceptually similar alternatives that have a shorter code representation, thereby survive the coding process and preserve the crispness of the original impression. 
\end{enumerate}

By simply manipulating the input data, we alter the way the fixed function codec behaves. We present evaluations across different datasets and codecs and show examples of how introducing GANs can lead to distinctly sharper looking images at low coding rates.

\section{Learning to Filter with a Surrogate Codec}
\begin{figure*}[ht]
	\centering
	\includegraphics[width=0.9\linewidth]{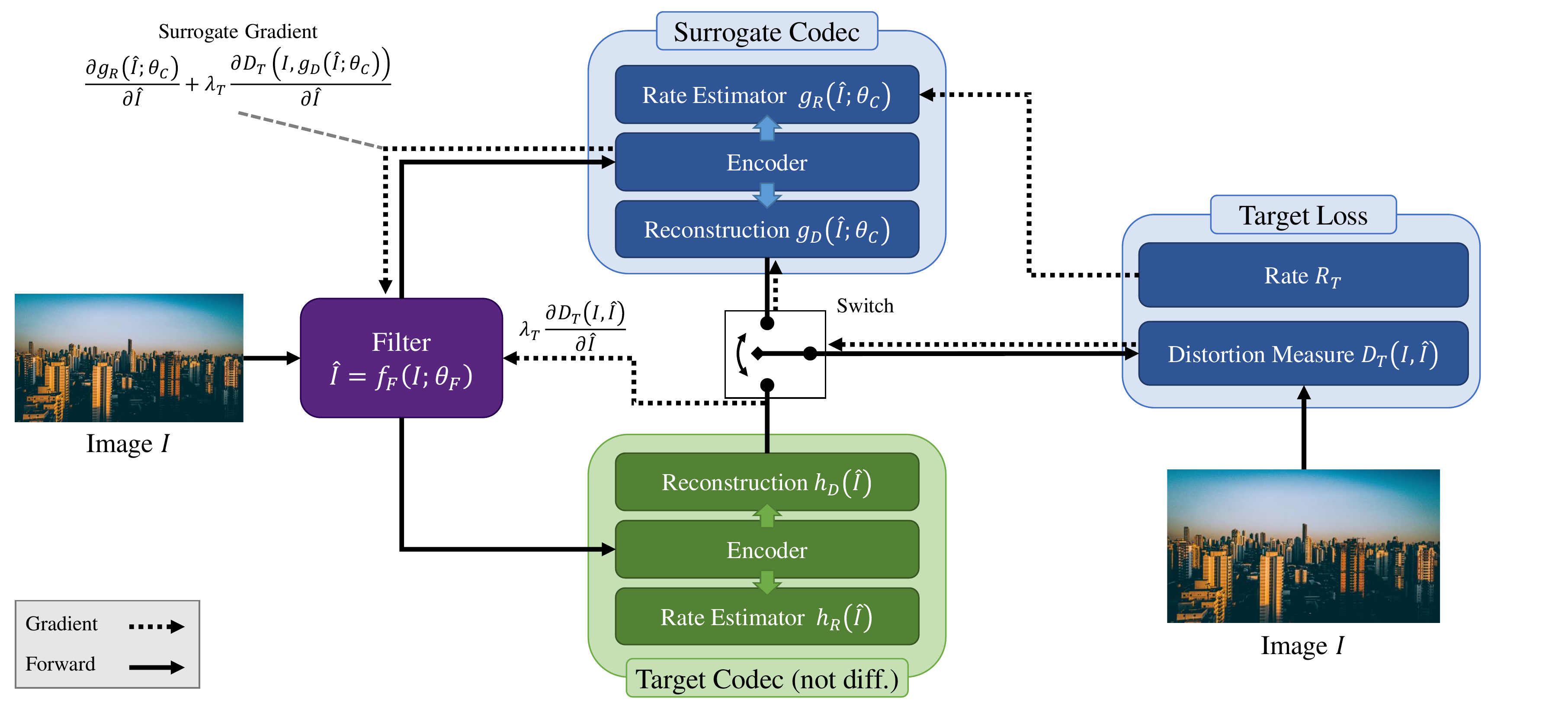}
	\caption{\textbf{Structural overview of our method.} The goal is to obtain a trained filter $f_F(I;\theta_F)$ to optimise the input image $I$ for encoding by a target codec. This target is typically not differentiable. A surrogate codec is used instead. It provides a differentiable rate estimate. For the reconstruction there are two options as indicated by the switch. The first is to take the surrogate's reconstruction, the second to invoke the target during the forward pass. The gradient flows back accordingly either through the surrogate's reconstruction or directly into the filter. The target distortion measure $D_T$ can be chosen freely. The surrogate codec is pre-trained with a distortion measure similar to the one of the target codec so as to imitate its behaviour. At testing time, the filter is applied to the image before it's encoded by the target codec.}
	\label{fig:structure}
\end{figure*}
\begin{table}[ht]
	\begin{tabular}{@{}lccccc@{}}
		\toprule
		Approach  & \multicolumn{2}{c}{Encoder} & & \multicolumn{2}{c}{Decoder} \\ \midrule
		& \begin{tabular}[c]{@{}l@{}}Remove\\Inform.\end{tabular} & \begin{tabular}[c]{@{}l@{}}Capture\\ Redund.\end{tabular} & Code & Rec. & \begin{tabular}[c]{@{}l@{}}Post-\\Filter\end{tabular} \\ \midrule
		Conv. Codec  &  \cmark   & \cmark  & \cmark  & \cmark  & \xmark/\cmark  \\
		Learn. Codec &  \cmark   & \cmark  & \cmark & \cmark  & \cmark \\
		Denoising    &    &  &      &      &  \cmark \\
		Inpainting   &    & \cmark & & & \cmark \\
		Choi et al. \cite{Choi2020} & \cmark   &   &     & \cmark &    \\
		Our's        & \cmark  &   &      &      &    \\ \bottomrule
	\end{tabular}
	\caption{Comparing different coding mechanism as to which part of a lossy coding pipeline they influence.}
	\label{tab:lpf_comparison}
\end{table}
Current coding mechanisms are most often instances of transform coding, where the data is first transformed, then quantised in the transform domain and finally encoded losslessly using a variant of arithmetic coding. Conventional codecs use a cosine- or sine-transform while learned codecs replace them with a convolutional neural network. This has the distinct advantage that the filters of the neural network can be adapted to leave out information that has low marginal coding gain, i.e. costly to encode and has only little benefit as measured by a given loss function. Table~\ref{tab:lpf_comparison} shows which functionality is invoked by different coding tools.

In codecs with fixed transforms this process needs to be manually emulated, which typically happens by varying the quantization granularity so that high frequency transform coefficients are quantised more coarsely than low frequency ones. Different rate-distortion trade-offs are then achieved by adjusting the granularity for all coefficients simultaneously. Omitting information happens implicitly through the quantization process. This, however, requires the transforms to separate information according to its importance so that varying quantisation granularity has the desired effect. The general underlying idea is that low frequency content is the most important and importance decreases towards higher frequencies. While for some content this adequately distinguishes noise from actual information, it does not hold in general. Hence, more adaptive filters are desirable, however difficult to adjust as their influence on the coding behaviour is unknown. 

Omitting unimportant information helps to significantly reduce the size of compressed multimedia data. For an optimal codec, it would be desirable to design the data filtering process with the distortion measure on the one hand and the transformation and coding procedure on the other in mind. Machine-learning based codecs can do this implicitly because all elements required are differentiable, hence their optimisation procedure allows the filters to adapt to the input data, the distortion measure, and the arithmetic coding procedure. For conventional codecs, this process is realized through extensive engineering and by relying on local optimisation of the code structure, for example how to partition the image into transform blocks or the choice of prediction modes, which makes it difficult to turn a codec into a differentiable function and obtain a gradient with respect to the input image.

In the machine learning literature this has been addressed as black box function optimisation, for example through the REINFORCE learning algorithm \cite{Williams1992} or the \textit{learning-to-learn} approaches \cite{Hochreiter2001,Andrychowicz2016,Chen2017}. These methods rely on an approximation of the unknown function to be optimised, though. Filtering of less important information is likely to require a very high precision of such an approximation, as the changes applied to the input image are expected to be small in magnitude. In preliminary experiments, we found that even a complex, deep model does not approximate a codec's rate and its decoded image accurately. This is most likely due to the complexity of the non-stationary error distribution that codecs produce, which varies between adjacent blocks. Choi et al.~\cite{Choi2020} present an approach for predicting optimised quantisation matrices for the JPEG codec. While they demonstrate effectiveness, their approach is limited to this (particularly simple) codec. Their approach also directly influences reconstruction (c.f. Table~\ref{tab:lpf_comparison}) because they are using a built-in codec tool, limiting the corrections their approach can perform.

In contrast, we present an approach that does not rely on approximating the codec directly but exploits a (differentiable) model that optimises the same objective, i.e. Eq.~\ref{eq:rate_distortion}. The structure of this approach is depicted in Figure~\ref{fig:structure}. Our objective is to approximate the gradient
\begin{equation}
\frac{\partial \mathcal{L}_h}{\partial I}=\frac{\partial h_R(I)}{\partial I}+\lambda\frac{\partial D(h_D(I),I)}{\partial I}
\end{equation}
of the loss function $\mathcal{L}_h$ of a codec $h$ with a rate $h_R(I)$, a decoded reconstruction $\tilde{I}=h_D(I)$ and a distortion measure $D(\hat{I},I)$ between the decoded and the original image $I$. Image compression models like \cite{Minnen2018} provide a differentiable codec $g$ with rate $g_R(I;\theta_C)$ and decoded reconstruction $g_D(I;\theta_C)$ both utilising some parameter vector $\theta_C$. Using this, an image $I$ can be manipulated by adding $\zeta^*$ where 
\begin{equation}
\label{eq:zeta_opt}
\zeta^*=\arg\min_{\zeta} \mathcal{L}_f=\arg\min_{\zeta} g_R\left(I+\zeta\right)+\lambda_T D_T\left(g_D\left(I+\zeta\right),I\right)
\end{equation}
can be optimised using a gradient descent based algorithm. The original codec then operates on the modified input image $\hat{I}=I+\zeta^*$ with the objective of achieving 
\begin{equation}
\label{eq:zeta_loss}
\mathcal{L}_h>\hat{\mathcal{L}}_h=h_R\left(\hat{I}\right)+\lambda_T D_T\left(h_D\left(\hat{I}\right), x\right)~.
\end{equation}
for a chosen target distortion $D_T$.

We introduce a filter $\hat{I}=f_F(I;\theta_F)$ to predict the filtered image $\hat{I}$ directly from the original image $I$. Its parameters $\theta_F$ can be trained using the gradient of the loss $\hat{\mathcal{L}}_h$ where the surrogate codec provides the rate estimation. The parameter $\lambda_T$ controls the trade-off between rate and distortion for the filter $f_F$.

Most codecs are designed for a particular \textit{intrinsic} distortion measure $D_I$. For common codecs like H265 or H264, coding tools and mechanisms were evaluated using the peak-signal-to-noise-ratio besides subjective viewing tests during standardization. This suggests that a codec may perform differently in terms of the rate-distortion trade-off when subjected to a different distortion measure. It is therefore likely that the measure used to optimise the parameters $\theta_C$ of the learned codec $g$ should be the same as that for which $h$ was designed. This will be confirmed by experiments. We will denote that measure by $D_S$ as it is the one to be approximated. When optimizing Equation~\ref{eq:zeta_opt} on the other hand, the distortion $D_T$ can be freely chosen so long as it is differentiable. 

With this framework in place, we are set to optimize the filter $f_F$ for a chosen particular target distortion $D_T$ while observing rate constraints as given by the surrogate codec's rate estimation. The following section briefly introduces the network architectures used in our experiments.
\subsection{Network Architectures}
\subsubsection{Filter}
\begin{figure*}[ht]
	\centering
	\includegraphics[width=0.9\linewidth]{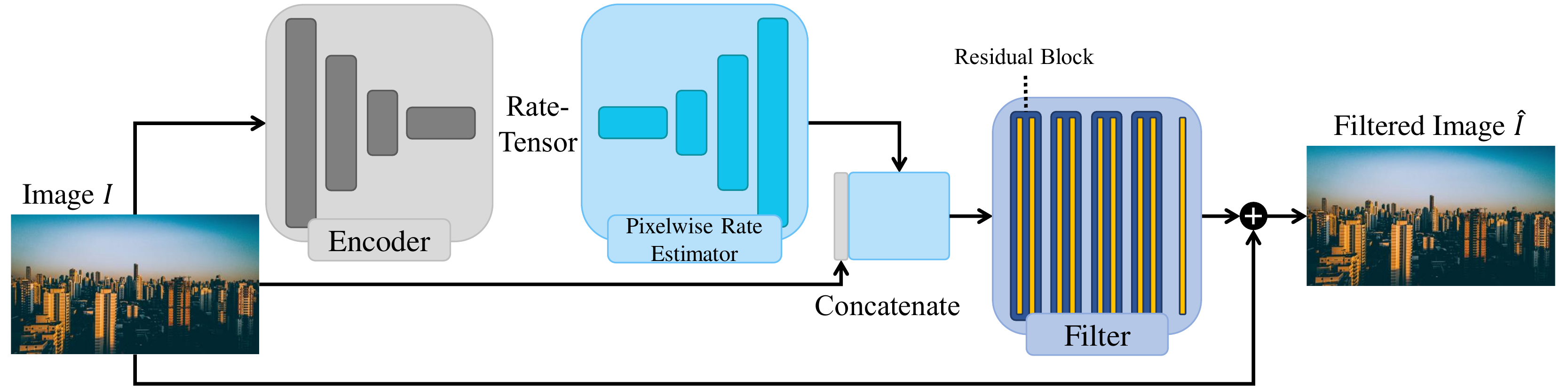}
	\caption{\textbf{Architecture of the filter component.} The encoder (gray) is taken from the surrogate codec, however, we do not use its latent representation but only its rate estimate, i.e. the function $h_R$. Its weights are fixed. The latent code's rate estimate is transformed into a pixel wise rate estimate via four successive transposed convolutions. The result is concatenated with the input image and fed to a filter consisting of four residual blocks.}
	\label{fig:filter}
\end{figure*}
The filter needs to remove information that is too costly to encode and has, according to the chosen distortion measure, little influence on the appearance. The chosen network architecture is depicted in Fig.~\ref{fig:filter}. The first stage is identical to the encoder and the rate estimation of the chosen surrogate codec. The general idea is that the filter first needs to estimate which regions of the image have a higher concentration of information. This is contained in the entropy estimates $H_{i,j,k}$. $H_{i,j,k}$ is of dimension $\frac{h}{16}\times\frac{w}{16}\times{C}$ where $C$ is the number of channels of the code tensor and $h$ and $w$ the dimensions of the image. For the surrogate model from \cite{Minnen2018} chosen here, this is $C=320$. The next step is to successively scale the code up to the size of the image using four layers of transposed convolutions with a stride two. We keep reducing the number of filters in each layer so as to have 16 output features per pixel. The idea behind not using a single value at each pixel is that we would like to represent a more complex distribution as the allocation of coding cost to single pixels is obviously challenging and a distribution can be interpreted as being conditional on the input image that is fused in the next step. The resulting pixel-wise entropy estimate $H_{\text{pixel}}\in\mathbb{R}^{h\times w\times 16}$ is concatenated with the input image $I$ to serve as input to the actual filter. The filter itself consists of four residual blocks followed by a convolutional layer at the end. We use 64 channels and kernels of size $3\times 3$ throughout the entire filter. 
\subsubsection{Discriminator}
We adopt a simple discriminator, as its only task is to recognize artefacts in the filtered image $\hat{I}$ and simple patterns in texture as opposed to larger structures or semantically coherent objects when images are generated from scratch.  Our discriminator has three consecutive stages to capture artefacts at different scales as previously described by \cite{Rippel2017}. Each stage consists of two residual blocks. Subsequent stages apply $2\times 2$ average pooling prior to the residual blocks. Each stage $s$ is followed by a single convolutional filter that computes the discriminator's prediction $p_s$. To fuse those predictions we first apply a sigmoid to convert them into probabilities before computing the mean over each stage:
\begin{equation}
\label{eq:gan}
\hat{y}_{\text{GAN}}(I)=\frac{1}{3}\sum_{s=1}^3\mathbb{E}\left[\text{sigmoid}\left(p_s(I)\right)\right]
\end{equation}
We noticed that only computing high level outputs at the discriminator does lead to low-level artefacts like blurriness in the image, an observation shared previously by Rippel and Bourdev \cite{Rippel2017}. Similar to \cite{Agustsson2018}, we used a least squares objective to train the discriminator, as previously introduced in \cite{Mao2017}. Discriminator and filter are optimized alternatingly, i.e. not using the same batch, for stability.
\section{Experiments}
\subsection{Experimental Conditions}
\subsubsection{Datasets} used in our evaluation are the low-resolution Kodak dataset (24 images, about 0.35MP) and the higher resolution "professional" validation dataset of the CLIC competition (41 images at more than 2MP). Our training data is sampled from the ImageNet dataset and CLIC's provided training data. We randomly resize an image so that its shorter side is between 512 and 1024 pixel in length and then sample a $256\times 256$px crop.
\subsubsection{Surrogate Model}
We use the deep convolutional neural network model proposed by Minnen et al. \cite{Minnen2018}. Besides a 4-layer encoder and decoder with generalized divisive normalization \cite{Balle2017} it uses context modelling to minimize the entropy of the code symbols. It has been shown to work well for both the MS-SSIM as well as the PSNR distortion metrics and achieve state-of-the-art performance on the Kodak test set. In addition, there is a reference implementation available \cite{Balle2018b,Begaint2020}. 
\subsubsection{Optimization}
Our models are implemented using PyTorch \cite{Paszke2017} and trained using the Adam \cite{Kingma2015} optimizer. A batch size of 8 is used with a learning rate of 1e-4 for 400.000 iterations, which is reduced to 1e-5 for another 100.000 iterations afterwards.
\subsubsection{Codecs} We use JPEG, WebP, BPG (H265/HEVC) \cite{Bellard} and the recently released AV1 to validate our hypothesis and demonstrate coding improvements. For all codecs we choose maximum quality over encoding speed and encode in the codec's native YCbCr colour space.
\subsection{Adaptation to MS-SSIM}
\label{sec:adapt_msssim}
\begin{figure}[t!]
	\centering
	\includegraphics[width=0.95\linewidth]{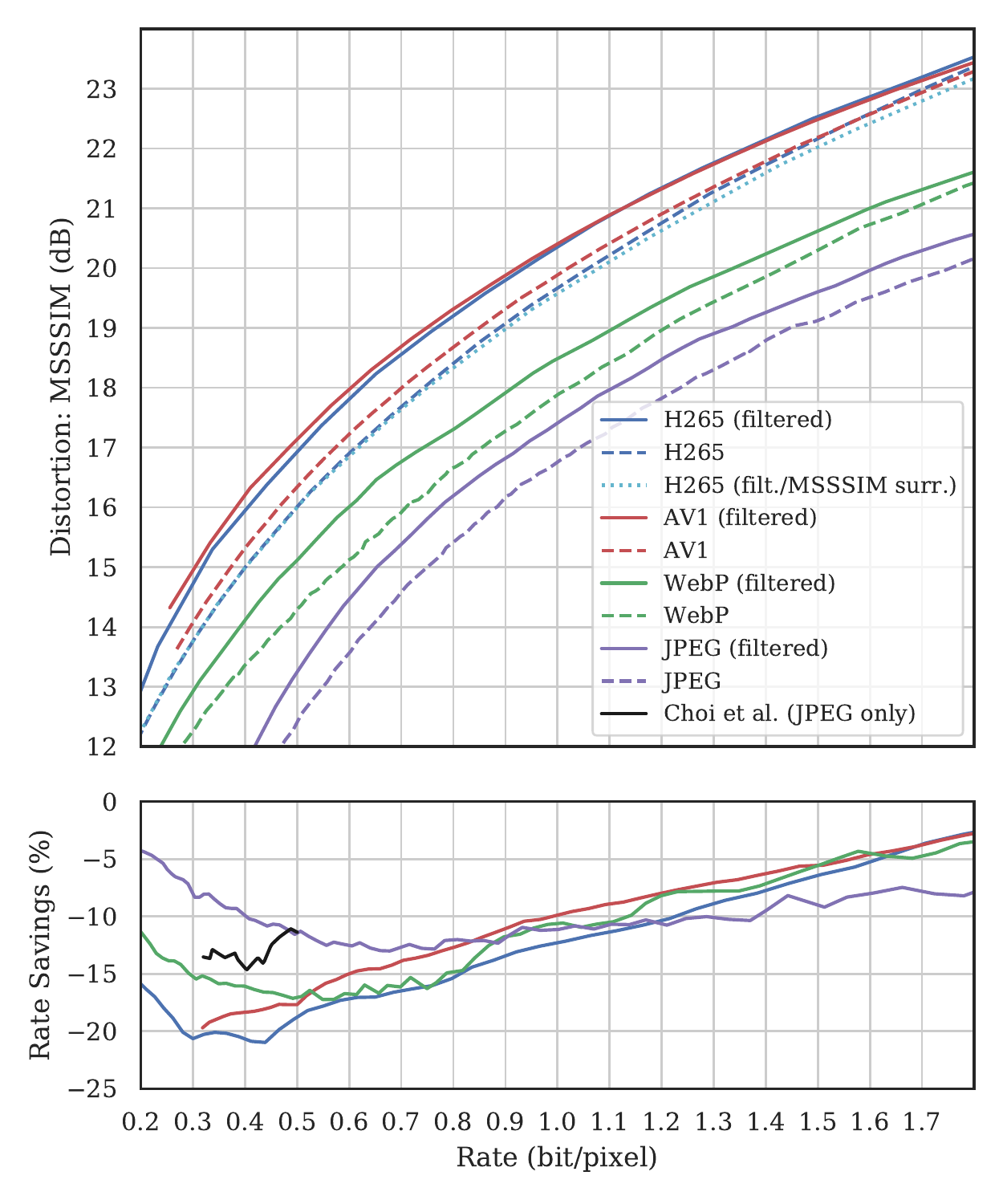}
	\caption{Rate-Distortion characteristic and relative rate savings for different codecs when surrogate-induced filtering is optimized for MS-SSIM and tested on the Kodak dataset. The dashed lines represent the rate-distortion characteristic of the codecs without filtering, the solid lines with filtering. The dotted line (cyan) is created using a surrogate codec trained on the MS-SSIM loss instead of PSNR. It shows no improvement over the original BPG (H265) codec as the surrogate distortion does not match the codec's. The bottom figure shows the rate savings over different rates, i.e. how much less rate the filtered codecs require relative to the original ones.}
	\label{fig:exp_msssim_kodak}
\end{figure}
We train a filter to minimise the rate while having only minimal impact on the MS-SSIM distortion measure. The idea is that most codecs were designed to minimise the mean squared error and also optimise that measure internally when choosing the best representation at encoding time. Not only can the mean squared error be computed faster, it also doesn't transcend block boundaries, making it a computationally suitable objective. Hence, "retargeting" a codec to MS-SSIM without having to change its encoder is appealing if it can be done by simply adding a filter at encoding time. 

We train three models at different rate-distortion trade-offs $\lambda_T$ to cover different parts of the rate range. The reason is that we found in preliminary experiments when optimising $\zeta^*$ directly, i.e. not training a filter $f_F(I;\theta_F)$, that the learned surrogate codecs perform well around their chosen operating point as determined by $\lambda_S$, however quickly diminish in performance once the trade-off point $\lambda$ is changed significantly. For this set of experiments, we chose $\lambda_S\in\left\lbrace0.1,0.2,0.4\right\rbrace$ for the MSE-optimized surrogate. When optimising Eq.~\ref{eq:zeta_loss} using the surrogate codec, the rate-distortion trade-off $\lambda_T$ for the target MS-SSIM loss needs to be chosen appropriately. We found $\lambda_T=500\lambda_S$ to work well by comparing at which pairs of $\lambda_{S,\text{MSE}},\lambda_{S,\text{MS-SSIM}}$ surrogate codecs optimized for each of the two distortions would yield similar rates.

There are limitations for very high bit rates as our retargeting process takes out information and hence the reconstruction error cannot be arbitrarily small. This is reflected in the graphs in Fig.~\ref{fig:exp_msssim_kodak}. Towards higher rates, where the error is very small, the effect of the filtering diminishes and fades with the filtered version eventually having a worse distortion than the original one. For the more advanced codecs AV1, BPG and WebP the rate savings peak for low rates, only for JPEG the rate savings are much less dynamic. The reason may lie in simplicity of the JPEG codec, there is little space for optimization as the coding and quantization process is static, only the quantization tables are optimized (over the entire image). Interestingly, for both datasets, the BPG codec initially performs worse than AV1, however, has higher adaptation coding gains so that the filtered variants of both codecs are almost on par.

\begin{figure}[]
	\centering
	\includegraphics[width=\linewidth]{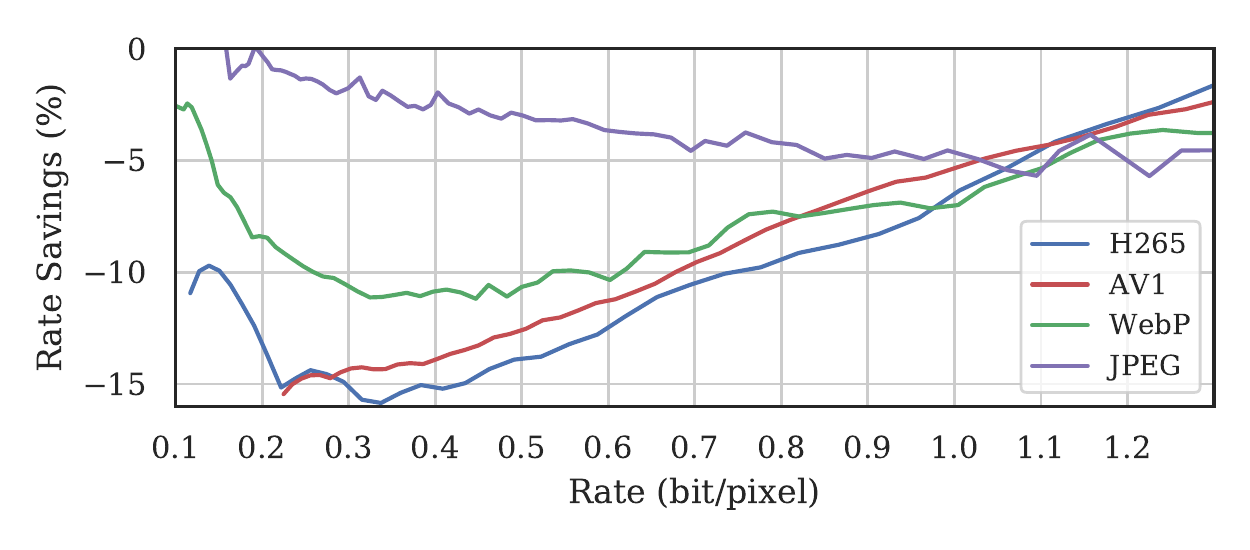}
	\caption{Relative rate savings for different codecs when surrogate-induced filtering is optimized for MS-SSIM and tested on the CLIC Professional (validation) dataset.}
	\label{fig:exp_msssim_prof_valid_rate_savings}
\end{figure}

Comparing the rate savings of the Kodak dataset in Fig.~\ref{fig:exp_msssim_kodak} to those of the validation part of the CLIC Professional dataset in Fig.~\ref{fig:exp_msssim_prof_valid_rate_savings},
one notices that the three advanced codecs show a similar behaviour in terms of rate savings. There's a peak early on followed by a steady, parallel decline. This indicates that at higher rates, the induced error from the filtering process is too high relative to the coding error. When optimizing only for numeric measures instead of perceptual ones, this limitation cannot be overcome. This is another reason to also look at perceptual losses like generative adversarial networks as discussed in Section~\ref{sec:gan}. Beyond this, the higher resolution images of the CLIC Professional dataset are more difficult to optimize yet still have coding gains up to 15\%. Higher resolutions often have a lower information density which makes it harder to filter out irrelevant parts.

Lastly, we also test with a surrogate codec optimised for $D_S=\text{MS-SSIM}$. This is shown as dotted, cyan coloured line in Fig.~\ref{fig:exp_msssim_kodak}. There is no improvement over the raw codec, showing that the surrogate's rate estimation needs to be based on the correct loss function to model a codec's rate allocation properly. 

\subsection{Generative Adversarial Network}
\label{sec:gan}
Generative Adversarial Networks have lead to remarkable results in various tasks where high resolution natural images are to be reconstructed such as super resolution \cite{Ledig2017} or learned image compression. In the context of adapting a codec, the GAN serves as an (additional) discrimination measure. 

Following \cite{Agustsson2018}, we employ mean squared error $\mathcal{L}_{\text{MSE}}$ and along with the VGG-based perceptual loss $\mathcal{L}_{\text{VGG}}$ proposed in \cite{Wang2018} in addition to the GAN loss $\mathcal{L}_{\text{GAN}}$ to ensure that the image content remains similar and only textural information changes. The final loss function for the filter adds these three distortion terms to the rate term $h_R$ of the surrogate codec:
\begin{equation}
\label{eq:loss_filter}
\mathcal{L}_{\text{Filter}}=\gamma_{\text{GAN}}\mathcal{L}_{\text{GAN}}+\gamma_{\text{VGG}}\mathcal{L}_{\text{VGG}}+\gamma_{\text{MSE}}\mathcal{L}_{\text{MSE}}+h_R
\end{equation}
where the different loss components can be weighted according to one's objectives.
The filter's GAN loss $\mathcal{L}_{\text{GAN}}(\hat{I})$ conditioned on the filtered image $\hat{I}=f_F(I;\theta_F)$ is given by the least squares objective from \cite{Mao2017}:
\begin{equation}
\mathcal{L}_{\text{GAN}}\left(I)\right)=\mathbb{E}_I\left[\left(1-y_\text{GAN}\left(f_F(I;\theta_F)\right)\right)^2\right]
\end{equation}
where $y_\text{GAN}\left(\hat{I}\right)$ is defined by Eq. \ref{eq:gan}. The discriminator's loss optimizes the opposite objective of $\mathcal{L}_{\text{GAN}}$:
\begin{equation}
\mathcal{L}_{\text{Discriminator}}=\mathbb{E}_I\left[\left(1-y_\text{GAN}\left(I\right)\right)^2\right]+\mathbb{E}_I\left[y_\text{GAN}\left(f_F(I;\theta_F)\right)^2\right]
\end{equation}
$\mathcal{L}_{\text{Discriminator}}$ and $\mathcal{L}_{\text{Filter}}$ are optimized in turn each other iteration. 

The choice of weights in Eq.~\ref{eq:loss_filter} determines what kind of sensitivity the GAN discriminator should apply. A lower weight $\gamma_{\text{GAN}}$ allows only for small changes in the image so that smaller artefacts are being corrected, however textures will not be exchanged for ones that have a shorter code length yet look more crisp when coded at low rates.

Setting the GAN's weight to $\gamma_{\text{GAN}}=5.0$ and $\gamma_{\text{VGG}}=0.01$, $\gamma_{\text{MSE}}=0.001$, the discriminator forces the filter to replace textures that are indistinguishably sharp yet have a shorter code length. The MSE and the VGG-based loss ensure that the texture is similar to the original. Examples of such a transformation is shown in Table~\ref{tab:lpf_gan_comparison}. Due to the low rate, the codec's aggressive quantization causes most details to wash out. At even lower coding rates, our pre-filtered images is able to preserve perceived details throughout the coding process. 

This way, without changing the mechanisms a conventional codec employs to represent an image, we can manipulate the codec through its input into optimizing for a rather perception-based objective. The benefit is that the shallow decoder that operates with only about 100 operations per pixel can be reused and the encoder is simply augmented and otherwise remains unchanged.

\begin{table*}
	\setlength\tabcolsep{1.5pt}
	\caption{Visual comparison, showing how our algorithm modifies certain textures within the image to mimic the original image's details with lower-code-length content as controlled by a generative adversarial network.}
	\begin{tabular}[c]{ccc}
		\toprule
		Original & H265 (unfiltered) & Ours (H265 filtered)\\
		\midrule
		\begin{tabular}[c]{@{}c@{}}
			\includegraphics[width=0.33\linewidth,trim={12 12 12 12},clip]{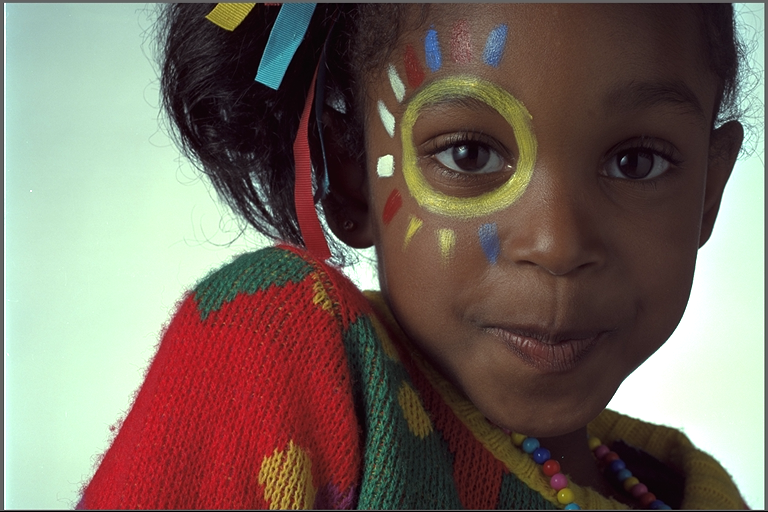} \\
			\includegraphics[width=0.25\linewidth]{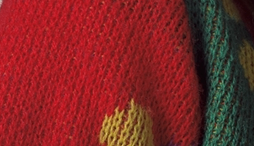} \\
			\quad
		\end{tabular}  &
		\begin{tabular}[c]{@{}c@{}}
			\includegraphics[width=0.33\linewidth,trim={12 12 12 12},clip]{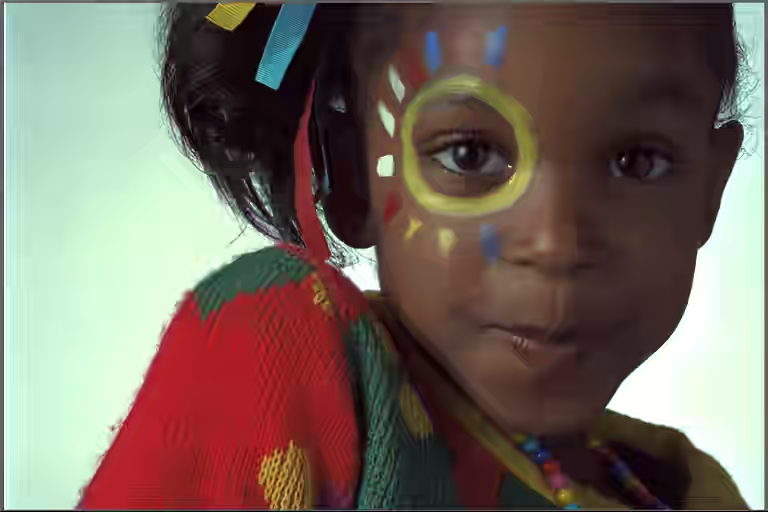} \\
			\includegraphics[width=0.25\linewidth]{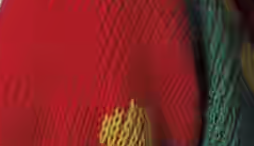} \\
			Rate: 0.086 bit/pixel			
		\end{tabular} &
		\begin{tabular}[c]{@{}c@{}}
			\includegraphics[width=0.33\linewidth,trim={12 12 12 12},clip]{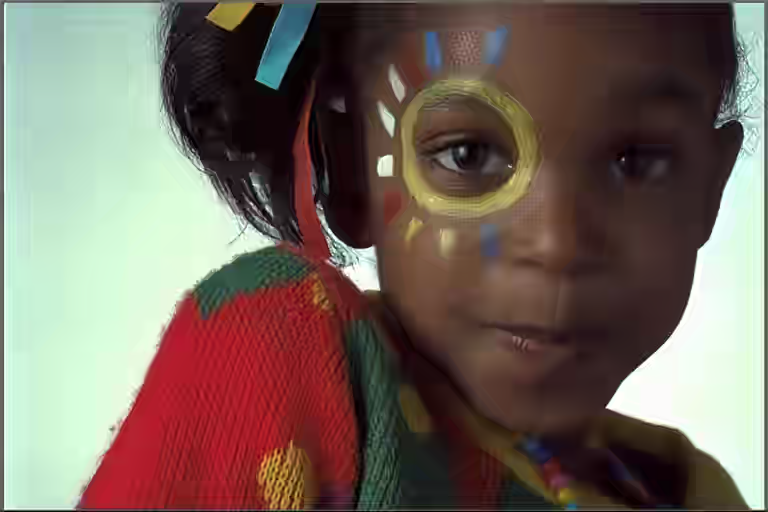} \\
			\includegraphics[width=0.25\linewidth]{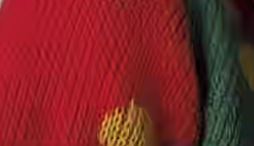} \\
			Rate: 0.084 bit/pixel		
		\end{tabular} \\		
		\begin{tabular}[c]{@{}c@{}}
			\includegraphics[width=0.33\linewidth,trim={12 12 12 12},clip]{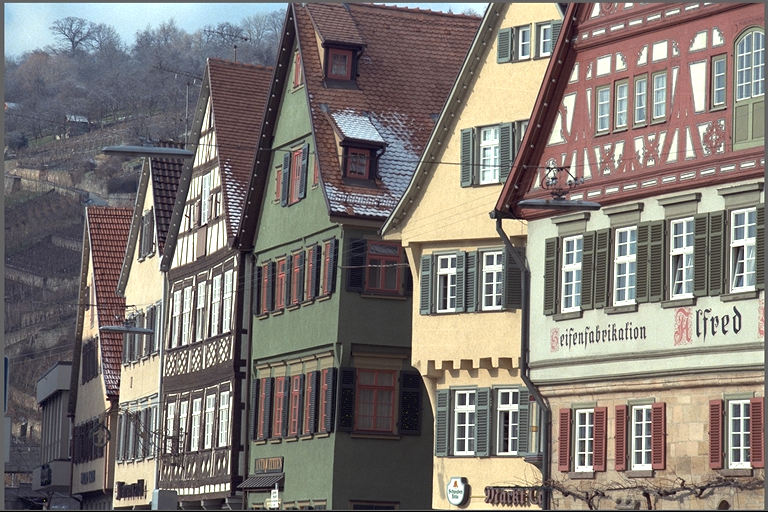} \\
			\includegraphics[width=0.25\linewidth]{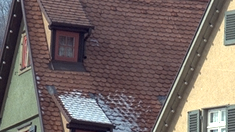} \\
			\quad
		\end{tabular}  &
		\begin{tabular}[c]{@{}c@{}}
			\includegraphics[width=0.33\linewidth,trim={12 12 12 12},clip]{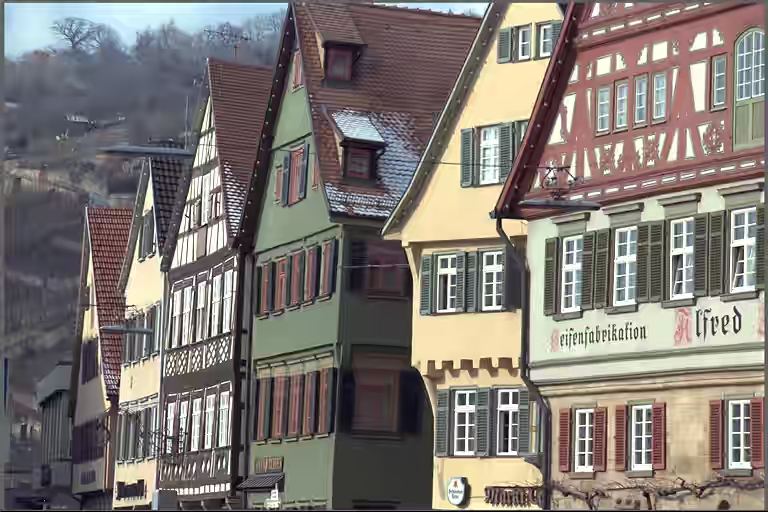} \\
			\includegraphics[width=0.25\linewidth]{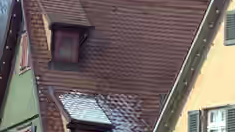} \\
			Rate: 0.045 bit/pixel			
		\end{tabular} &
		\begin{tabular}[c]{@{}c@{}}
			\includegraphics[width=0.33\linewidth,trim={12 12 12 12},clip]{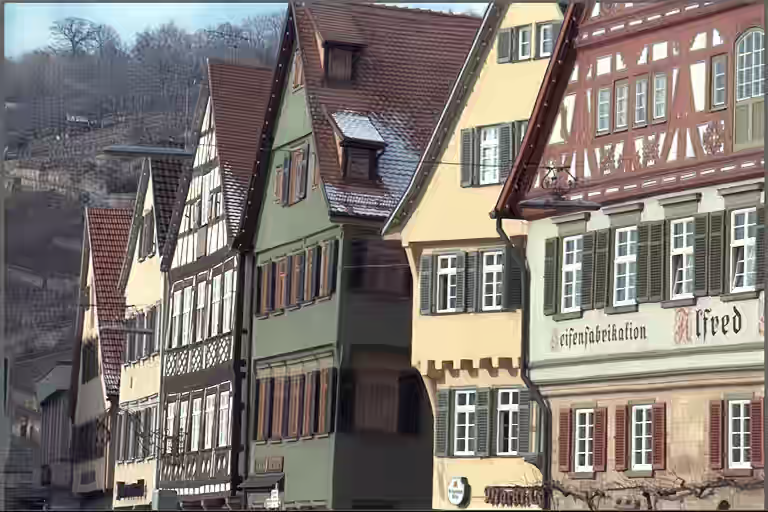} \\
			\includegraphics[width=0.25\linewidth]{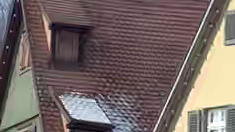} \\
			Rate: 0.042 bit/pixel		
		\end{tabular} \\
		\begin{tabular}[c]{@{}c@{}}
			\includegraphics[width=0.33\linewidth,trim={12 12 12 12},clip]{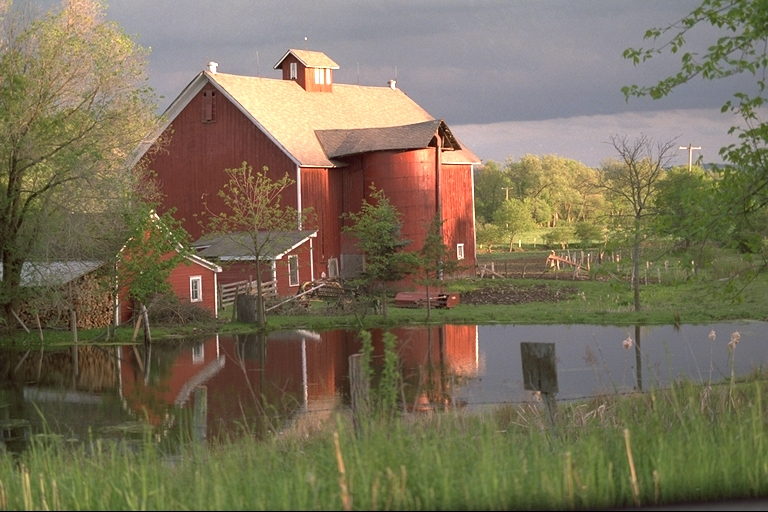} \\
			\includegraphics[width=0.25\linewidth]{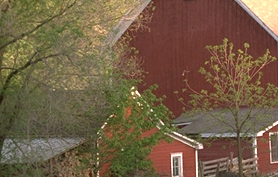} \\
			\quad
		\end{tabular}  &
		\begin{tabular}[c]{@{}c@{}}
			\includegraphics[width=0.33\linewidth,trim={12 12 12 12},clip]{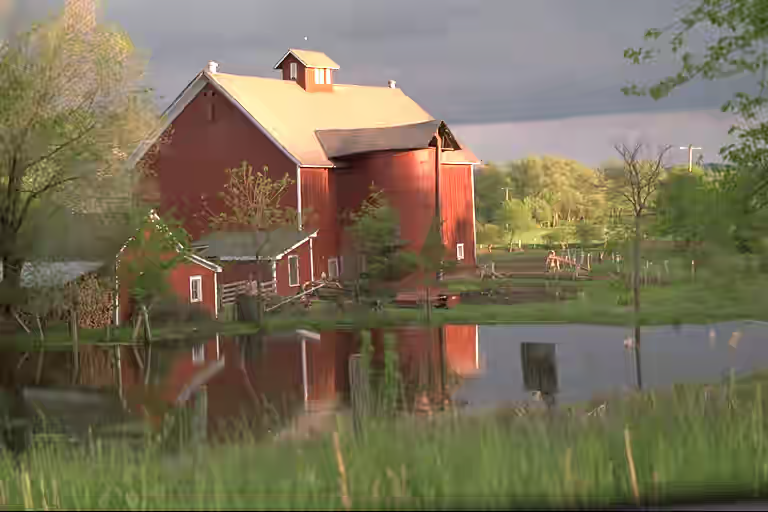} \\
			\includegraphics[width=0.25\linewidth]{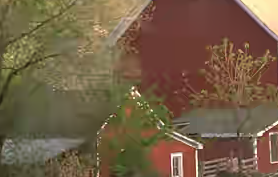} \\
			Rate: 0.190 bit/pixel			
		\end{tabular} &
		\begin{tabular}[c]{@{}c@{}}
			\includegraphics[width=0.33\linewidth,trim={12 12 12 12},clip]{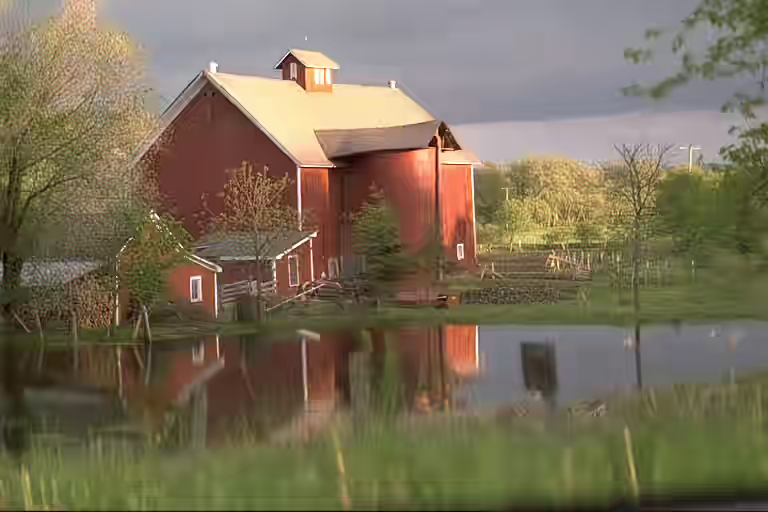} \\
			\includegraphics[width=0.25\linewidth]{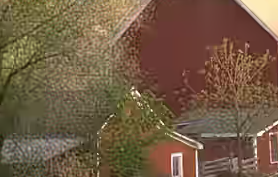} \\
			Rate: 0.192 bit/pixel		
		\end{tabular} \\
		\bottomrule
	\end{tabular}
	\label{tab:lpf_gan_comparison}
\end{table*}

\subsection{Task-Aware Compression}
Task-aware compression takes a non-perceptual task into account when compressing data. In this work we follow \cite{Choi2020} and use the ImageNet \cite{JiaDeng2009} object recognition task. To learn the filter, we use the pretrained ResNet-18 \cite{He2016a} available from PyTorch. For this task, the filter receives the classifier's feedback on which features are important. This necessitates two changes to our algorithm. 

The first is to initialise the filter with a very low variance so that the generated correction signal has a very low amplitude at the beginning of the training. If one omits this, the (initially random) correction signal still gets adjusted to strengthen certain features, however, the learned codec's rate estimate doesn't transfer well. Apparently, the reason for this is that there are certain high frequency signals that are cheap for a learned but expensive for a conventional codec and the filter tends to exploit those because they are closer to its starting point, i.e. the high variance random initialization. 

The second measure is to use the target codec's decoded version of the filtered image $\hat{I}$, $h_D(\hat{I})$, to obtain feedback from the classifier about which features are important. While this slows training down due to $B$ calls to the conventional codec during the forward pass for batch size $B$, it resolves the same issue as in the previous paragraph. The information that goes missing are likely to be high frequency signals involved in rather fine-grained recognition. If the surrogate codec's decoded image, $g_D(\hat{I})$, was used directly, the filter may overfit and exploit high-frequency low-cost signals. 

\begin{figure}
	\centering
	\includegraphics[width=0.9\linewidth]{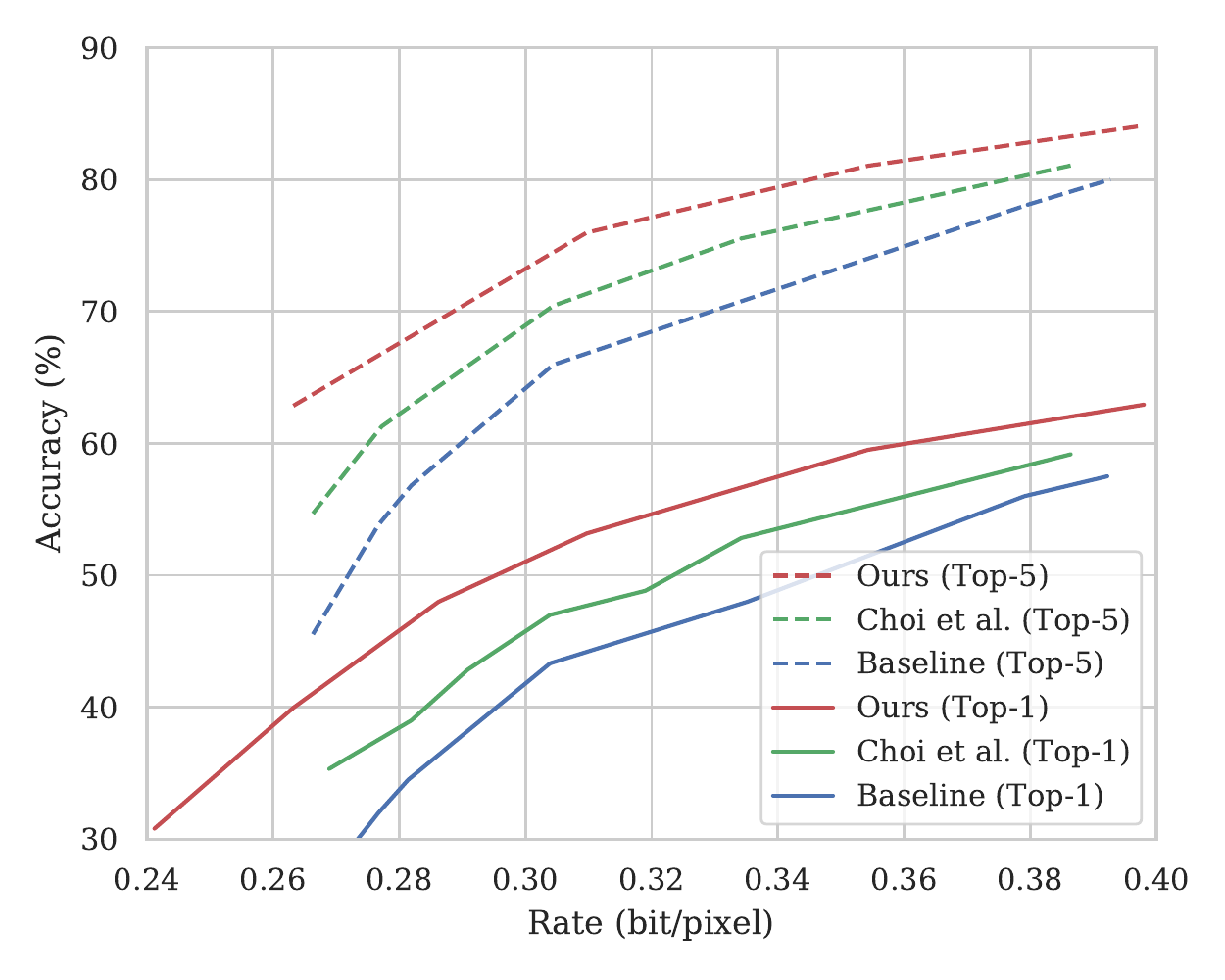}
	\caption{Comparison of our work to Choi et al. (\cite{Choi2020}) on the ImageNet image recognition task with Inception-v3. Baselines from \cite{Choi2020}.}
	\label{fig:lpf_exp_imagenet_choi}
\end{figure}

For testing, in accordance with \cite{Choi2020}, we use a pretrained Inception-v3 architecture \cite{Szegedy2016}. We train a single model using JPEG at a quality level of 30 at training time. For testing, we filter the images once and then encode at different quality levels. The results are shown in Fig.~\ref{fig:lpf_exp_imagenet_choi}. Although \cite{Choi2020} has a more complicated model that mimics the JPEG codec, our approach outperforms them across all rates. One reason is that \cite{Choi2020} can only manipulate quantisation tables while we can actually make arbitrary changes to the image. In this, our method, though simpler, is more powerful.

\section{Related Work}
Choi et al.'s work \cite{Choi2020} is related to ours in that they improve a conventional codec as well. Their method, however, is limited to the JPEG codec by explicit modelling which is difficult to extend to more complex codecs. In the experiments section we showed that our approach performs similarly to them when retargeting a codec to another metric and performs favourably on task-specific compression.

Our work is related to other approaches that employ generative adversarial networks to reconstruction tasks, for image compression this are the recent works \cite{Agustsson2018,Mentzer2020}. Different from our approach, their work treats the problem from a conceptual perspective, ignoring the complexity constraints or the benefit of using existing coding schemes, by employing many-layered neural networks in the image formation process.

Similar to their work, we employ techniques from learned image compression. Recursive architectures have been proposed \cite{Toderici2015,Toderici2016,Johnston2017} as well as adoptions of variational auto-encoders \cite{Balle2017,Li2017,Baig2017,Agustsson2017,Rippel2017} with various extensions to the latent code model \cite{Balle2019,Liu2019,Klopp2018,Minnen2018,Balle2018,Mentzer2018}. Within a few years, these techniques have closed the gap to conventional codecs, however, largely ignore complexity bounds typically posed on their conventional counterparts with decoders requiring three to four orders of magnitude more operations per pixel. Again, this is an important difference to our approach, as we target an existing coding pipeline. Furthermore, those works have previously been extended into the video coding domain, typically by adding and encoding optical flow \cite{Lu2019,Habibian2019,Rippel2018}.

Other works have improved upon existing codecs by adding neural networks as denoising modules. Pretrained denoisers such as \cite{Jia2019,Zhang2018} are effective but like the aforementioned approaches do add heavily to the decoder complexity. Using online adaptation of a small neural network and signalling its parameters as part of the code, \cite{Klopp2020} showed that the complexity could be reduced significantly while still delivering a coding gain similar to that of pretrained denoising networks. On the downside, these methods require changes to the codec architecture and target only numerical losses, i.e. optimizing Equation \ref{eq:rate_distortion}, no perceptual ones. 

\section{Conclusion}
We have demonstrated how machine learning based image compression algorithms can be repurposed as surrogate gradient generators to train image filters to alter the coding characteristic of conventional codecs without explicitly modelling their behaviour. Experiments demonstrate retargeting to the MS-SSIM distortion measure leads to over 20\% coding gain, an advantage over a codec-specific optimisation approach on task-aware image compression, and how adding a generative adversarial network can enable preservation of sharp textures even in very low rate conditions.

Our approach opens up a number of possible directions for future research. Modelling existing or upcoming codecs to a higher accuracy could enable even better results. Traditional, hand-optimized image signal processing pipelines as they are found in smart phones for example could be optimized end-to-end from acquisition to image formation and to coding. The introduced concepts extend to the motion picture domain as well, which could enable codec refitting beyond still images.

Finally, the proposed technique does not require changing the coding pipeline. Encoder and decoder stay in place. Hence, in contrast to many other image coding improvements, it can utilize existing efficient hardware implementations of conventional decoders.
{\small
\bibliographystyle{ieee_fullname}
\bibliography{main}
}

\end{document}